\documentclass[pra,twocolumn,amsmath,showpacs]{revtex4}
\usepackage{graphicx,bm}
\begin{document}

\hyphenation{mol-e-cul-es}

\author{J. Aldegunde}
\email{E-mail: Jesus.Aldegunde@durham.ac.uk}

\affiliation{Department of Chemistry, Durham University, South
Road, DH1 3LE, United Kingdom}

\author{Jeremy M. Hutson}
\email{E-mail: J.M.Hutson@durham.ac.uk}

\affiliation{Department of Chemistry, Durham University, South
Road, DH1 3LE, United Kingdom}

\title{The hyperfine energy levels of alkali metal dimers: \\
ground-state homonuclear molecules in magnetic fields}

\date{\today}

\begin{abstract}
We investigate the hyperfine energy levels and Zeeman splittings for
homonuclear alkali-metal dimers in low-lying rotational and
vibrational states, which are important for experiments designed to
produce quantum gases of deeply bound molecules. We carry out
density-functional theory (DFT) calculations of the nuclear
hyperfine coupling constants. For nonrotating states, the zero-field
splittings are determined almost entirely by the scalar nuclear
spin-spin coupling constant. By contrast with the heteronuclear
case, the total nuclear spin remains a good quantum number in a
magnetic field. We also investigate levels with rotational quantum
number $N=1$, which have long-range anisotropic
quadrupole-quadrupole interactions and may be collisionally stable.
For these states the splitting is dominated by nuclear quadrupole
coupling for most of the alkali-metal dimers and the Zeeman
splittings are considerably more complicated.
\end{abstract}



%
\maketitle
\section{Introduction}
\label{intro} There have been enormous advances over the last year
in experimental methods to produce ultracold molecules in their
rovibrational ground state at microkelvin temperatures. Ospelkaus
{\em et al.}\ \cite{Ospelkaus:2008} produced KRb molecules in
high-lying states by magnetoassociation (Feshbach resonance tuning)
and then transferred them by stimulated Raman adiabatic passage to
levels of the $^1\Sigma^+$ ground state bound by more than 10 GHz.
This was then extended by Ni {\em et al.}\ \cite{Ni:KRb:2008} to
produce molecules in $(v,N)=(0,0)$, where $v$ and $N$ are the
quantum numbers for molecular vibration and mechanical rotation.
Danzl {\em et al.}\ \cite{Danzl:v73:2008, Danzl:ground:2008} have
carried out analogous experiments on Cs dimers, while Lang {\em et
al.}\ \cite{Lang:cruising:2008, Lang:ground:2008} have produced
Rb$_2$ molecules in the lowest rovibrational level of the lowest
triplet state. There have also been considerable successes in direct
photoassociation to produce low-lying states \cite{Sage:2005,
Hudson:PRL:2008, Viteau:2008, Deiglmayr:2008}.

A major goal of the experimental work is to produce a stable
molecular quantum gas. However, such a gas can form only if (i)
a large number of molecules are in the same hyperfine state and
(ii) the molecules are stable to collisions that occur in the
gas. In particular, inelastic collisions that transfer internal
energy into relative translational energy cause heating and/or
trap loss. It is thus very important to understand the
hyperfine structure of the low-lying levels and its dependence
on applied electric and magnetic fields. In a previous paper,
we explored the hyperfine levels of heteronuclear alkali metal
dimers in rotationless levels \cite{Aldegunde:polar:2008}. The
purpose of the present paper is to extend this work to
homonuclear molecules, which have important special features.
We also explore $N=1$ levels, which may be collisionally stable
for homonuclear molecules and which interact with longer-range
forces than $N=0$ levels.

\section{Molecular Hamiltonian}
\label{sec:MolHam} The Hamiltonian of a diatomic molecule in
the presence of an external magnetic field can be decomposed
into five different contributions: the electronic, vibrational,
rotational, hyperfine and Zeeman terms. For $^{1}{\rm \Sigma}$
molecules in a fixed vibrational level, the first two terms
take a constant value and the rotational, hyperfine and Zeeman
parts of the Hamiltonian may be written
\cite{Ramsey:1952,Brown,Bryce:2003}
\begin{equation}
\label{Htot} H = H_{\rm rot} + H_{\rm hf} + H_{\rm Z},
\end{equation}
where
\begin{eqnarray}
 \label{eq:Hrot} H_{\rm rot}  &=& B_v\bm{N}^{2}-D_v\bm{N}^{2}\cdot\bm{N}^{2}; \\
  \nonumber \label{eq:Hhf}H_{\rm hf}  &=& H_{\rm Q} + H_{\rm IN} + H_{\rm
  t} + H_{\rm sc} = \sum_{i=1}^{2}\bm{V}_{i}:\bm{Q}_{i} \\
  & + & \sum_{i=1}^{2} c_{i}
\,\bm{N}\cdot\bm{I}_{i} +c_3\,\bm{I}_{1}\cdot\bm{T}\cdot\bm{I}_2
+c_4\,\bm{I}_{1}\cdot\bm{I}_2;\\
 \label{eq:Hz} H_{\rm Z} &=& -g_{\rm r}\mu_{\rm N}
\,\bm{N}\cdot\bm{B}
 - \sum_{i=1}^{2}g_{i}\mu_{\rm N}
\,\bm{I}_{i}\cdot\bm{B} (1-\sigma_{i}).
\end{eqnarray}
where the index $i$ refers to each of the nuclei in the
molecule. $\bm{N}$, $\bm{I}_{1}$ and $\bm{I}_2$ represent the
operators for mechanical rotation and for the spins of nuclei 1
and 2. The rotational and centrifugal constants of the molecule
are given by $B_v$ and $D_v$ (but centrifugal distortion is
neglected in the present work). We use $\bm{N}$ rather than
$\bm{J}$ for mechanical rotation because we wish to reserve
$\bm{J}$ for the angular momentum including electron spin for
future work on triplet states.

The hyperfine Hamiltonian of equation \ref{eq:Hhf} consists of four
different contributions. The first is the electric quadrupole
interaction $H_{\rm Q}$, with coupling constants $(eqQ)_{1}$ and
$(eqQ)_2$. It represents the interaction of the nuclear quadrupoles
($eQ_{i}$) with the electric field gradients $q_i$ created by the
electrons at the nuclear positions. The second is the spin-rotation
term $H_{\rm IN}$, which describes the interaction of the nuclear
magnetic moments with the magnetic moment created by the rotation of
the molecule. Its coupling constants are $c_1$ and $c_2$. For a
homonuclear molecule with identical nuclei, $(eqQ)_{1}=(eqQ)_2$ and
$c_1=c_2$. The last two terms represent the interaction between the
two nuclear spins; there is both a tensor component $H_{\rm t}$,
with coupling constant $c_3$, and a scalar component $H_{\rm sc}$,
with coupling constant $c_4$. The second-rank tensor $\bm{T}$
represents the angular part of a dipole-dipole interaction.

The Zeeman Hamiltonian $H_{\rm Z}$ has both rotational and
nuclear Zeeman contributions characterized by $g$-factors
$g_{\rm r}$, $g_1$ and $g_2$. For homonuclear molecules
$g_{1}=g_2$. The nuclear shielding tensor ${\bm \sigma}_{i}$ is
approximated here by its isotropic part $\sigma_i$; terms
involving the anisotropy of ${\bm \sigma}_{i}$ are extremely
small for the states considered here.

The nuclear $g$-factors and the quadrupole moments of the
nuclei are experimentally known \cite{Stone:2005}.

For homonuclear molecules we neglect the effect of electric
fields, though in principle there are small effects due to
anisotropic polarizabilities and the molecular quadrupole
moments can interact with the gradients of inhomogeneous
fields.

\section{Evaluation of the coupling constants}
\label{sec:eval} The rotational $g$-factors are known
experimentally for all the homonuclear alkali metal dimers
\cite{Brooks:1963}. However, the only such species for which
the nuclear hyperfine coupling constants have been determined
accurately is Na$_2$ \cite{Esbroeck:1985}. We have therefore
evaluated the remaining coupling constants using
density-functional theory (DFT) calculations performed with the
Amsterdam density functional (ADF) package \cite{ADF1,ADF3}
with all-electron basis sets and including relativistic
corrections. A full description of the basis sets, functionals,
etc.\ used in the calculations has been given in our previous
paper on heteronuclear systems \cite{Aldegunde:polar:2008}. In
the present work, the calculations were carried out at the
equilibrium geometries ($R_{e}=2.67$ \AA\ for Li$_2$
\cite{Hessel:1979}, 3.08 \AA\ for Na$_{2}$ \cite{Kusch:1978},
3.92 \AA\ for K$_2$ \cite{Engelke:1984}, 4.21 \AA\ for Rb$_2$
\cite{Amiot:1985} and 4.65 \AA\ for Cs$_2$ \cite{Raab:1982}).
This give results that are approximately valid not only for
$v=0$ states but also for other low-lying vibrational states.

The values for the coupling constants are given in table
\ref{tb:hd}. It may be seen that the DFT results for Na$_2$ are
within about 30\% of the experimental values, and similar
accuracy was obtained for other test cases in our previous work
\cite{Aldegunde:polar:2008}. The accuracy is likely to be
comparable for the other cases studied here. This level of
accuracy is adequate for the purpose of the present paper,
which aims to explore the qualitative nature of the Zeeman
patterns. Most of our conclusions are insensitive to the exact
magnitudes of the coupling constants.

\begin{table*}[bhtp]
\caption{Nuclear quadrupole moment ($Q$), electric quadrupole
coupling constant ($eqQ$), nuclear $g$-factor ($g$),
spin-rotation coupling constant ($c_1$), tensor spin-spin
coupling constant ($c_3$), scalar spin-spin coupling constant
($c_4$), absolute value of the $c_4/(eqQ)$ ratio, isotropic
part of the nuclear shielding ($\sigma$) and rotational
$g$-factor ($g_{\rm{r}}$) for the homonuclear alkali dimers.
All the quantities except the nuclear quadrupole moments, the
nuclear \emph{g}-factors and the rotational \emph{g}-factors
were evaluated using DFT calculations (see section
\ref{sec:eval}). Both experimental \cite{Esbroeck:1985} and
theoretical results are presented for Na$_{2}$.} \label{tb:hd}
\begin{tabular}{cccccccccc}
\hline\noalign{\smallskip} & $Q$(fm$^{2}$) & $eqQ$(MHz) & $g$ & $c_1$(Hz) & $c_3$(Hz) & $c_4$(Hz) & $|c_4/(eqQ)|$&
$\sigma$(ppm) & $g_{\rm{r}}$ \\
\noalign{\smallskip}\hline\noalign{\smallskip}
$^6$Li$_2$ & -0.082 & 0.00123 & 0.822 & 161 & 137 & 32 & 0.026 & 102 &  0.1259\\
$^7$Li$_2$ & -4.06 & 0.0608 & 2.171 & 365 & 955 & 226 & 0.0037 & 102 &  0.1080\\
$^{23}$Na$_2$ (Exp.) & 10.45 & -0.459 & 1.479 & 243 & 303 &  1067 & 0.0023 & --- & 0.0386 \\
$^{23}$Na$_2$ (DFT) & --- & -0.456 & --- & 299 & 298 &  1358 & 0.0030 & 613 & --- \\
$^{39}$K$_2$ & 5.85 & -0.290 & 0.261 & 35 & 5 & 106 & 0.00036 & 1313 & 0.0212 \\
$^{40}$K$_2$ & -7.3 & 0.362 & -0.324 & -42 & 8 & 163 & 0.00045 & 1313 & 0.0207 \\
$^{41}$K$_2$ & 7.11 & -0.353 & 0.143 & 18 & 2 & 32 & 0.000091 & 1313 & 0.0202\\
$^{85}$Rb$_2$ & 27.7 & -2.457 & 0.541 & 63 & 30 & 2177 & 0.00089 & 3489 & 0.0095 \\
$^{87}$Rb$_2$ & 13.4 & -1.188 & 1.834 & 209 & 346 & 25021 & 0.021 & 3489 & 0.0093 \\
$^{133}{\rm Cs}_2$ & -0.355 & 0.0486 & 0.738 & 96 &  119 & 12993 & 0.27 & 6461 & 0.0054 \\
\noalign{\smallskip}\hline
\end{tabular}
\end{table*}

\section{Hyperfine energy levels} \label{sec:Molenlev}
Our previous work \cite{Aldegunde:polar:2008} showed that the
zero-field splitting for heteronuclear diatomic molecules in
$N=0$ states is determined almost entirely by the scalar
nuclear spin-spin interaction. This remains true for
homonuclear molecules in $N=0$ states. We show below that for
$N>0$ the electric quadrupole interaction is dominant for all
the homonuclear dimers except Cs$_2$ and $^{6}$Li$_2$, with
smaller but significant contributions from the remaining
coupling constants.

For all systems except Li$_2$, the scalar spin-spin coupling is
considerably stronger than the spin-rotation and tensor
spin-spin couplings. Knowledge of the nuclear quadrupole
coupling constant $eQq$ and the scalar spin-spin coupling
constant $c_4$ is therefore sufficient to understand the
hyperfine splitting patterns. We will focus here on
$^{85}$Rb$_2$ and $^{87}$Rb$_2$, which form a convenient pair
that approximately cover the range of values of the ratio
$|c_4/(eqQ)|$. Lang {\em et al.}\ \cite{Lang:ground:2008} have
produced $^{87}$Rb$_2$ in the lowest rovibrational level of the
lowest triplet state, but as far as we are aware not yet in the
singlet state.

The hyperfine energy levels are obtained by diagonalizing the
matrix representation of the Hamiltonian (\ref{Htot}) in a
basis set of angular momentum functions. In order to facilitate
the assignment of quantum numbers to the energy levels, two
different basis sets are employed,
\begin{eqnarray}
  \label{eq:cb} |(I_1 I_1) I M_{ I} N M_{ N}\rangle & & (\mbox{spin-coupled basis}); \\
  \label{eq:tcb} |(I_1 I_1) I  N F M_{ F}\rangle & & (\mbox{fully coupled basis}).
\end{eqnarray}
where $I$ and $F$ are the total nuclear spin and total angular
momentum quantum numbers and $M_I$ and $M_{F}$ are their
projections onto the $Z$ axis defined by the external field.
The matrix elements of the different terms in the Hamiltonian
in each of the basis sets are calculated through standard
angular momentum techniques \cite{Zare}. Explicit expressions
are given in the Appendix.

For homonuclear molecules, nuclear exchange symmetry dictates that
not all possible values of the total nuclear spin $I$ can exist for
each rotational level. For molecules in $^1\Sigma^+$ states, only
even $I$ values can exist for even $N$ and only odd $I$ for odd $N$.
This is true for either fermionic or bosonic nuclei but is reversed
for $^3\Sigma^+$ states. Table \ref{tb:Ival} summarizes the $I$-$N$
pairs compatible with the antisymmetry of the wave function under
nuclear exchange for the Rb$_2$ isotopomers.

\begin{table}
\caption{Values of $I$ permitted by the nuclear exchange
symmetry for even and odd rotational levels of $^{85}$Rb$_2$
and $^{87}$Rb$_2$.}
\label{tb:Ival}       
\begin{tabular}{ccccc}
\hline\noalign{\smallskip}
& & $N$ & & $I$  \\
\noalign{\smallskip}\hline\noalign{\smallskip}
 & & even & & 0, 2, 4 \\
\raisebox{2ex}{$^{85}$Rb$_2$($I_{\rm{Rb}}=5/2$)} & & odd & & 1, 3, 5 \\[2ex]
 & & even & & 0, 2 \\
\raisebox{2ex}{$^{87}$Rb$_2$($I_{\rm{Rb}}=3/2$)} & & odd & & 1, 3 \\
\noalign{\smallskip}\hline
\end{tabular}
\end{table}

\subsection{Zeeman splitting for $N=0$ homonuclear alkali dimers}
\begin{figure}[t]
\includegraphics[width=0.95\linewidth]{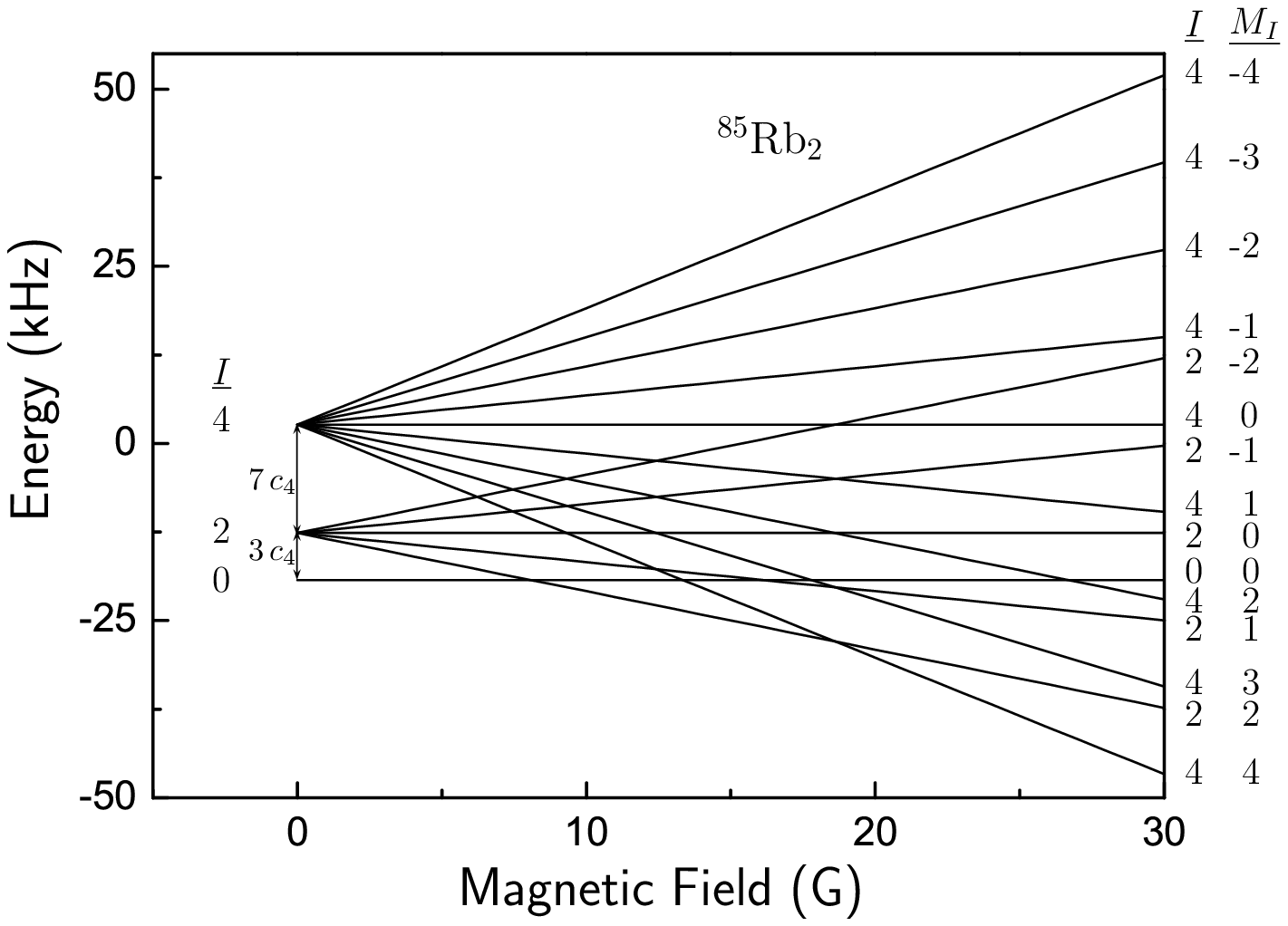}
\includegraphics[width=0.95\linewidth]{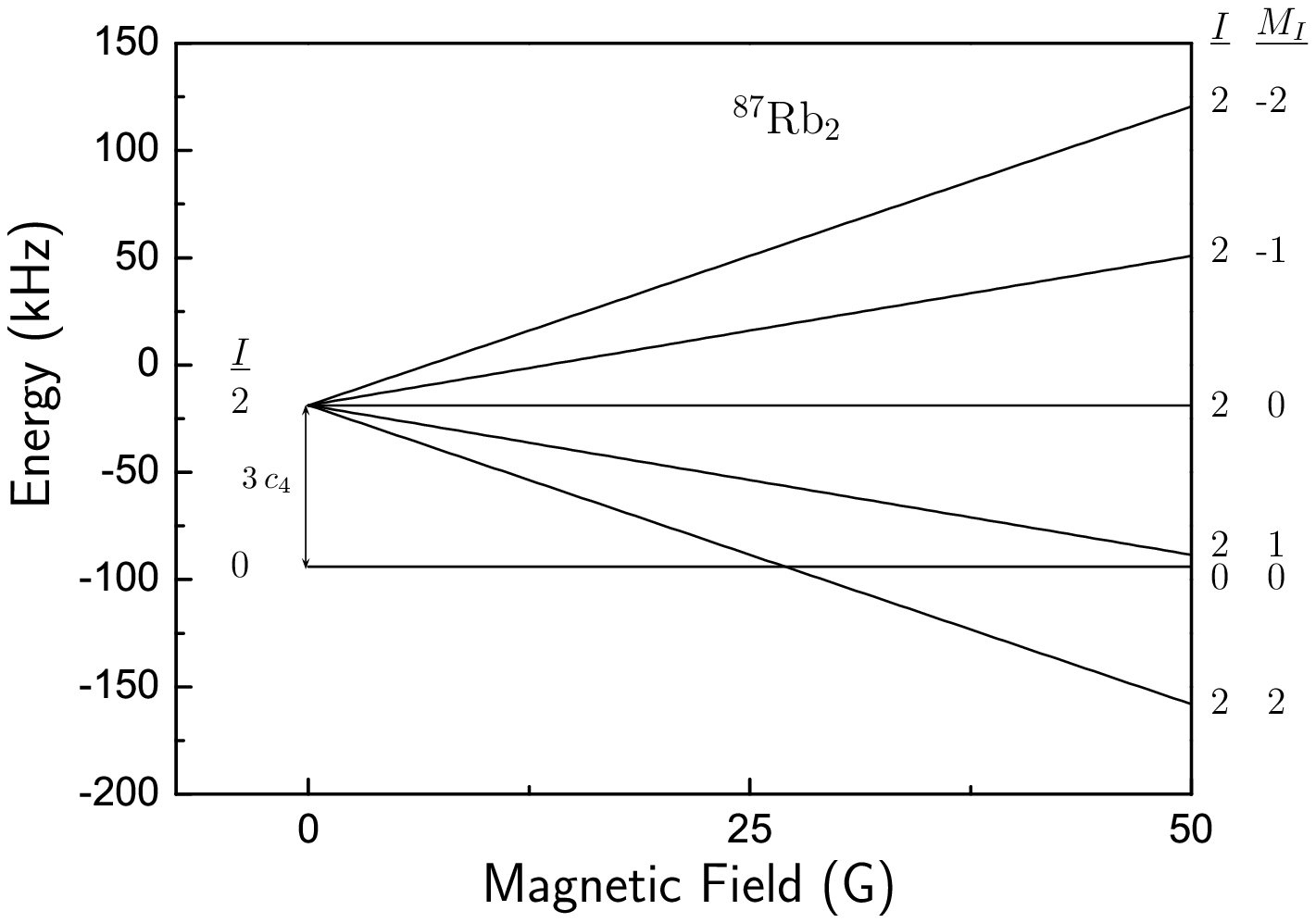}
\caption{\label{fig:01}%
Zeeman splitting of hyperfine levels for $N=0$ states of
$^{85}$Rb$_2$ (upper panel) and $^{87}$Rb$_2$ (lower panel).}
\end{figure}

The Zeeman splittings for the $N=0$ hyperfine levels of
$^{85}$Rb$_2$ and $^{87}$Rb$_2$ are shown in figure
\ref{fig:01}. The zero-field splittings are in most respects
similar to those found for heteronuclear molecules in the
ground rotational state \cite{Aldegunde:polar:2008}. The
similarities can be summarized as follows:
\begin{itemize}
\item The scalar nuclear spin-spin interaction and the
    nuclear Zeeman effect are the only two terms in the
    molecular Hamiltonian with nonzero diagonal elements
    for $N=0$.

\item The electric quadrupole and the tensor nuclear
    spin-spin interactions are not diagonal in $N$,
    coupling the $N$, $N+2$ and $N-2$ rotational levels.
    This means that the energy levels should be converged
    by including in the calculations as many rotational
    levels as necessary. However, the coupling constants
    $eQq$ and $c_3$ are very much smaller than the
    rotational spacings, so that in practice it is adequate
    to include one excited rotational level. Convergence
    for $N=0$ is reached with $N_{\rm max}=2$ and
    convergence for $N=1$ is reached with $N_{\rm max}=3$.
\item The scalar spin-spin interaction is diagonal in both
    the spin-coupled and fully coupled basis sets, which
    for $N=0$ are identical,
  \begin{eqnarray}
&& \langle (N=0) I M_I|
c_4\,\bm{I}_{1}\cdot\bm{I}_2| (N=0) I
M_{I}
   \rangle  \nonumber \\ &&\qquad =
\label{eq:Hescj0} \frac{1}{2} c_4 [I(I+1)-2I_{\rm Rb}(I_{\rm
Rb}+1)].
\end{eqnarray}
Except for a very small contribution coming from the
coupling with $N=2$ levels, these diagonal elements
determine the zero-field splitting.
\end{itemize}

%
%

Despite the similarity of the zero-field levels, there are
important differences between the Zeeman splittings for
heteronuclear and homonuclear molecules. For heteronuclear
dimers \cite{Aldegunde:polar:2008}, levels with the same $M_I$
but different $I$ exhibit avoided crossings as a function of
magnetic field. Because of this, $I$ is no longer a good
quantum number at high field but the individual nuclear spin
projections $M_{I1}$ and $M_{I_2}$ become nearly conserved. For
homonuclear dimers, however, different energy levels that
correspond to the same value of $M_{I}$ are parallel, so that
no avoided crossings appear as a function of the field. Both
$I$ and $M_{I}$ remain good quantum numbers regardless of the
value of the magnetic field but $M_{I1}$ and $M_{I_2}$ are not
individually conserved. This is illustrated in figure
\ref{fig:01}. It arises because the nuclear Zeeman term, which
is the only nondiagonal term for $N=0$ in the heteronuclear
case, is diagonal for homonuclear molecules. Its nonzero
elements are given by equation \ref{eq:HIZ1} of the Appendix,
\begin{eqnarray}
&&\langle (N=0) I M_{I}| H_{\rm IZ} |
(N=0) I M_{I}
   \rangle \nonumber \\
\label{eq:Hnz0} &&\qquad = -g_{\rm Rb}\mu_{\rm N}B_{\rm Z}(1-\sigma_{\rm
Rb})M_{I}.
\end{eqnarray}
The nuclear Zeeman term is diagonal because the $g$-factors of
the two nuclei are equal and not because of nuclear exchange
symmetry. The $N=0$ block of the molecular Hamiltonian for a
heteronuclear dimer with two identical nuclear $g$-factors
would also be diagonal.

The conservation of the total nuclear spin $I$ and
non-conservation of $M_{I1}$ and $M_{I2}$ at high fields may
have important consequences for the selection rules in
spectroscopic transitions used to produce ultracold molecules
and for the collisional stability of molecules in excited
hyperfine states.

\subsection{Zeeman splitting for $N=1$ homonuclear alkali dimers}
Ultracold homonuclear molecules in $N=1$ states are
particularly interesting because they are likely to be stable
with respect to inelastic collisions to produce $N=0$, at least
for collisions with non-magnetic species such as other
molecules in $^1\Sigma$ states. Such collisions cannot change
the nuclear spin symmetry and thus cannot change $N$ from odd
to even. Inelastic collisions may well be stronger for
collisions of molecules in triplet states, because of magnetic
interactions between electron and nuclear spins. Transitions
between odd and even rotational levels are permitted in
atom-exchange collisions, such as occur in collisions with
alkali metal atoms \cite{Soldan:2002, Cvitas:bosefermi:2005,
Cvitas:hetero:2005, Quemener:2005, Cvitas:li3:2007,
Hutson:IRPC:2007}.

Homonuclear molecules do not possess electric dipole moments
but do have quadrupole moments. The quadrupole-quadrupole
interaction is anisotropic and is proportional to $R^{-5}$, so
is longer-range than the $R^{-6}$ dispersion interaction that
acts between neutral atoms and molecules. The
quadrupole-quadrupole interaction averages to zero for
rotationless states ($N=0$), but not for $N>0$. Quantum gases
of rotating homonuclear molecules may thus exhibit anisotropic
effects.

For $N>0$, all the terms in the Hamiltonian (\ref{Htot}) have
matrix elements diagonal in $N$. Some of these are nondiagonal
in hyperfine quantum numbers, so the energy level patterns are
much more intricate. The zero-field splitting is dominated in
most cases by the electric quadrupole interaction and the
scalar nuclear spin-spin term. The remaining constants ($c_1$
and $c_3$) make much smaller contributions except for the two
Li$_{2}$ isotopomers. For $^{6}$Li$_{2}$, all the terms in the
hyperfine hamiltonian contribute significantly. For
$^{7}$Li$_{2}$, the splitting is dominated by the electric
quadrupole interaction but contributions from all the remaining
terms are significant.

\begin{figure}[t]
\includegraphics[width=0.95\linewidth]{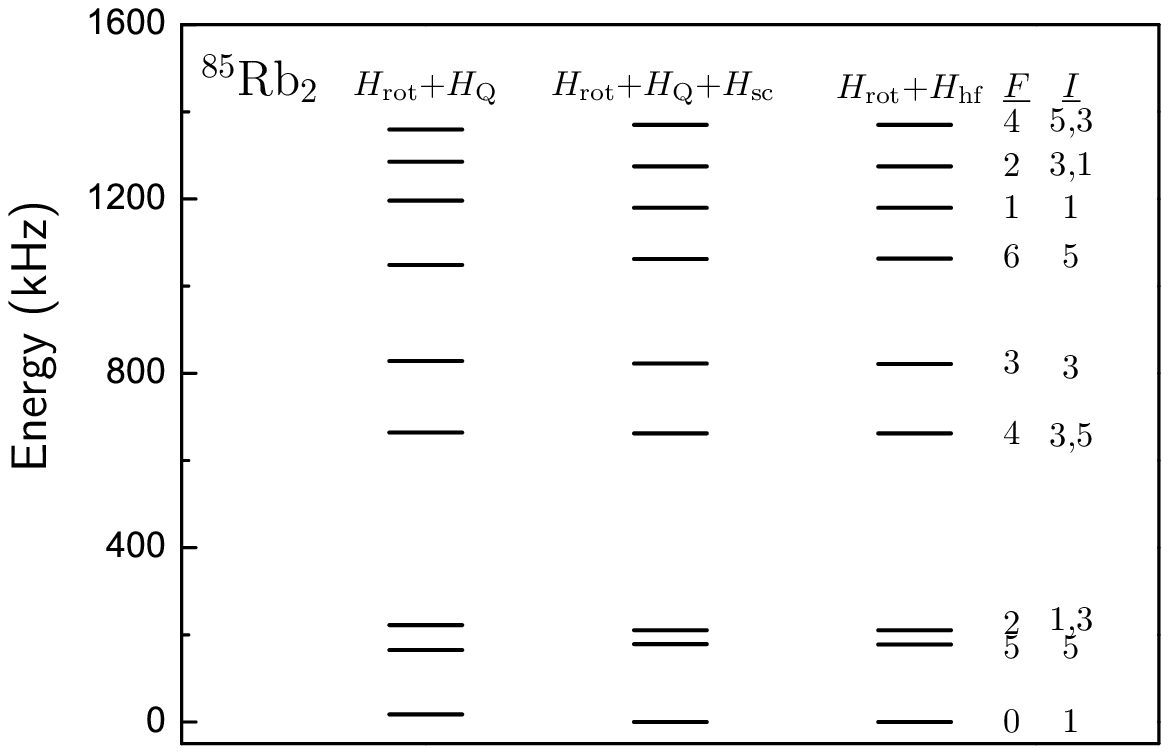}
\vspace{0.5cm}
\includegraphics[width=0.95\linewidth]{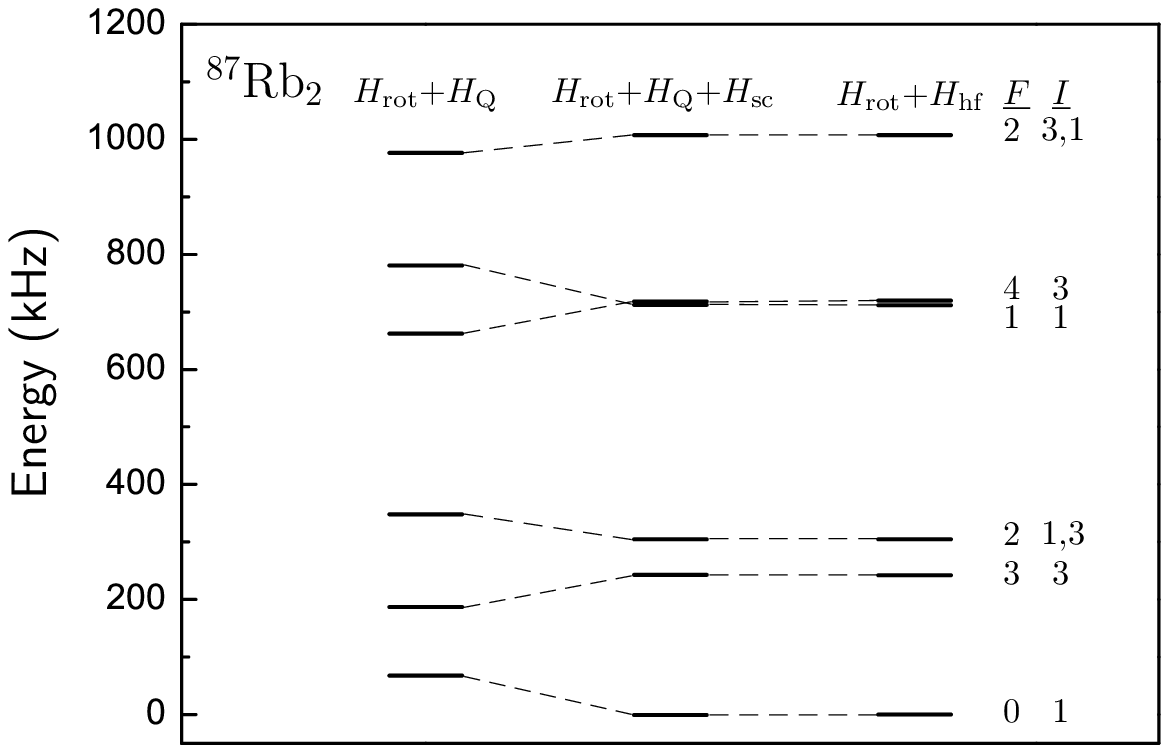}
\caption{\label{fig:02}%
Zero-field hyperfine splitting for $N=1$ states of
$^{85}$Rb$_2$ (upper panel) and $^{87}$Rb$_2$ (lower panel).
For each species, the hyperfine energy levels obtained when
only the rotational and electric quadrupole terms are included
are show in the left column. The effect of adding the scalar
spin-spin interaction is displayed in the central column and,
finally, the right column shows the splitting when the whole
hyperfine Hamiltonian is considered. All the energies are
referred to the lowest hyperfine level for the complete
Hamiltonian.}
\end{figure}

Figure \ref{fig:02} shows the ``building-up" of the zero-field
$N=1$ hyperfine energy levels for $^{85}$Rb$_2$ and
$^{87}$Rb$_2$ in three steps: first, only the rotational and
the electric quadrupole terms are considered; secondly, the
scalar spin-spin interaction is included; and thirdly, the
spin-rotation and the tensor spin-spin interaction terms are
added to complete the hyperfine Hamiltonian. For $^{85}$Rb$_2$,
the electric quadrupole term alone determines the energy level
pattern, while for $^{87}$Rb$_2$ there is a significant
additional contribution from the scalar spin-spin interaction,
attributable to the relatively large value of $c_4$ for this
molecule (see table \ref{tb:hd}).

%
%
%
%
\begin{figure}[t]
\includegraphics[width=0.95\linewidth]{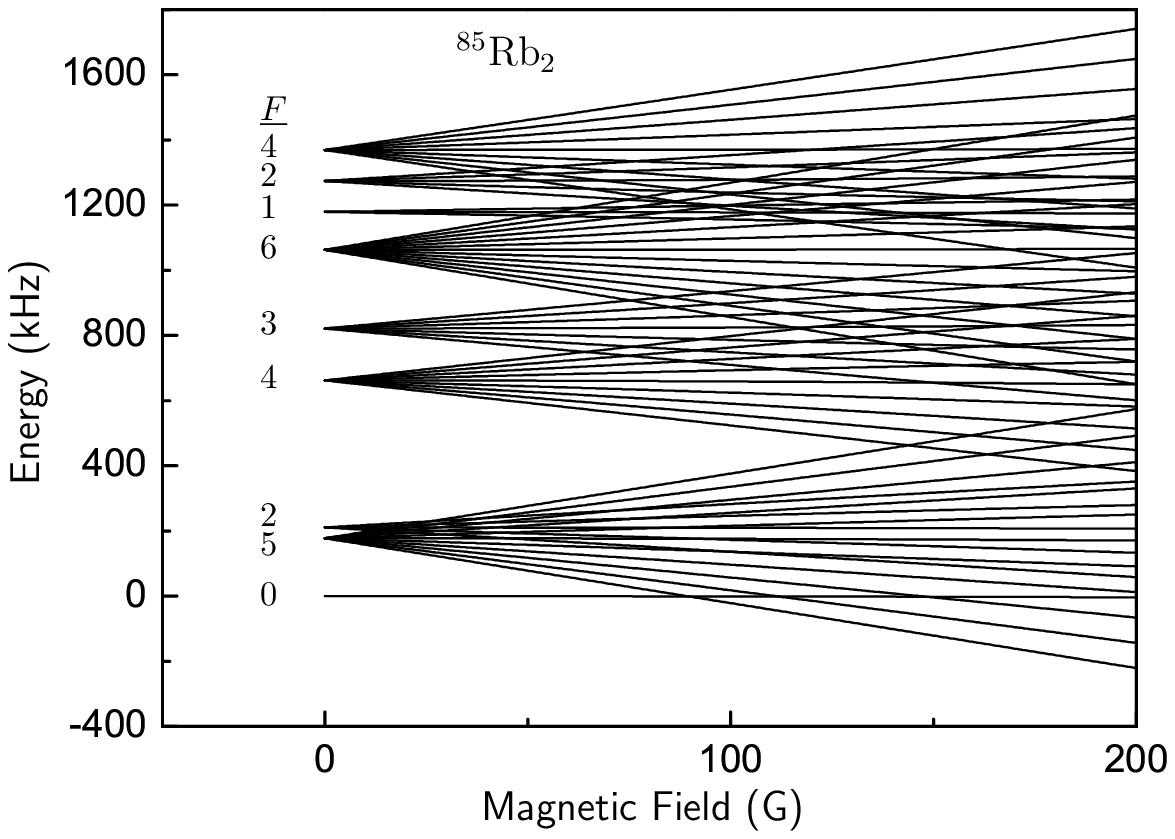}
\includegraphics[width=0.95\linewidth]{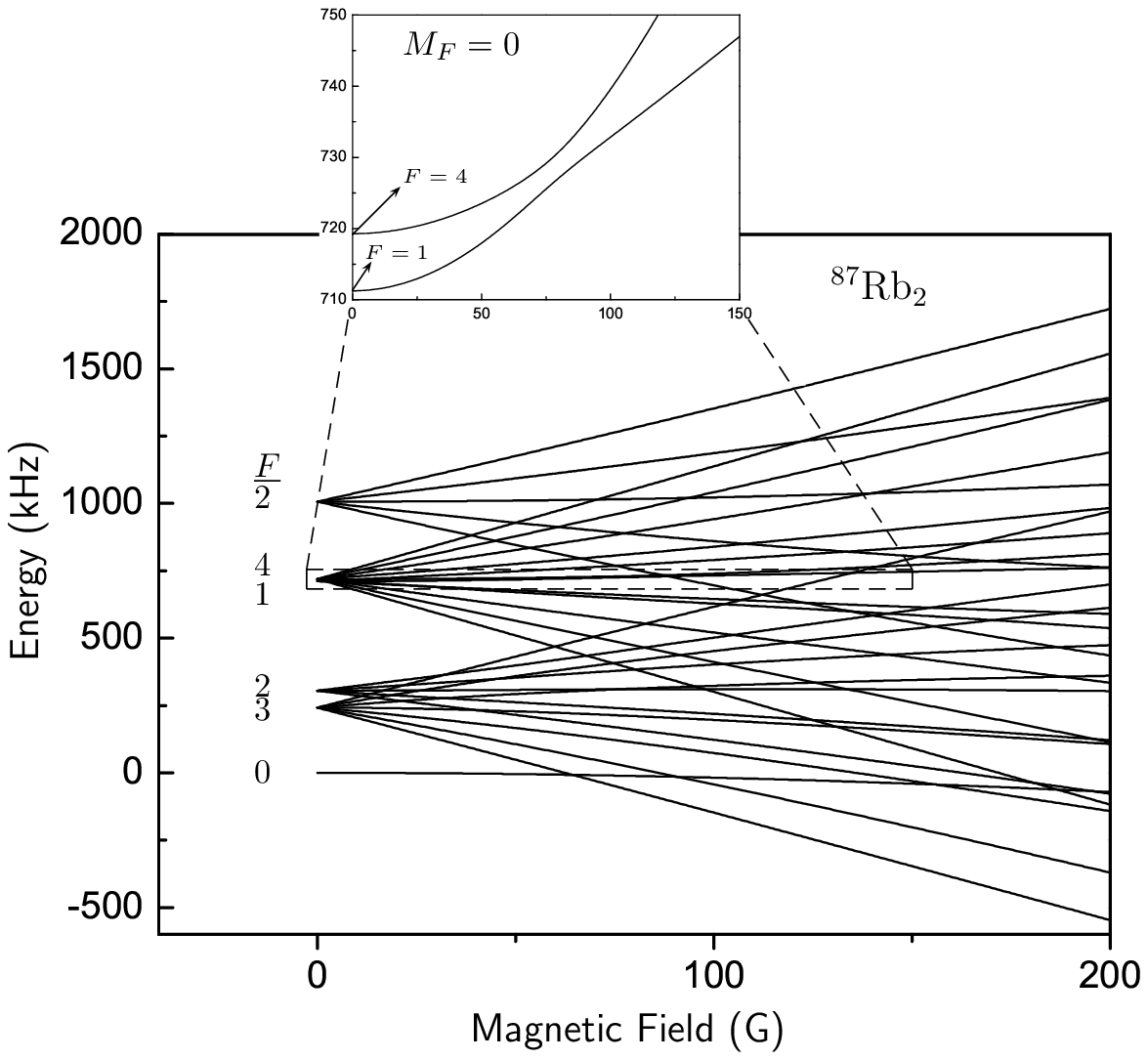}
\caption{\label{fig:03}%
Zeeman splitting for $N=1$ states of $^{85}$Rb$_2$ (upper
panel) and $^{87}$Rb$_2$ (lower panel). The inset for
$^{87}$Rb$_2$ shows the avoided crossing of $M_F=0$ states.}
\end{figure}
%
%
\begin{figure}[t]
\includegraphics[width=0.95\linewidth]{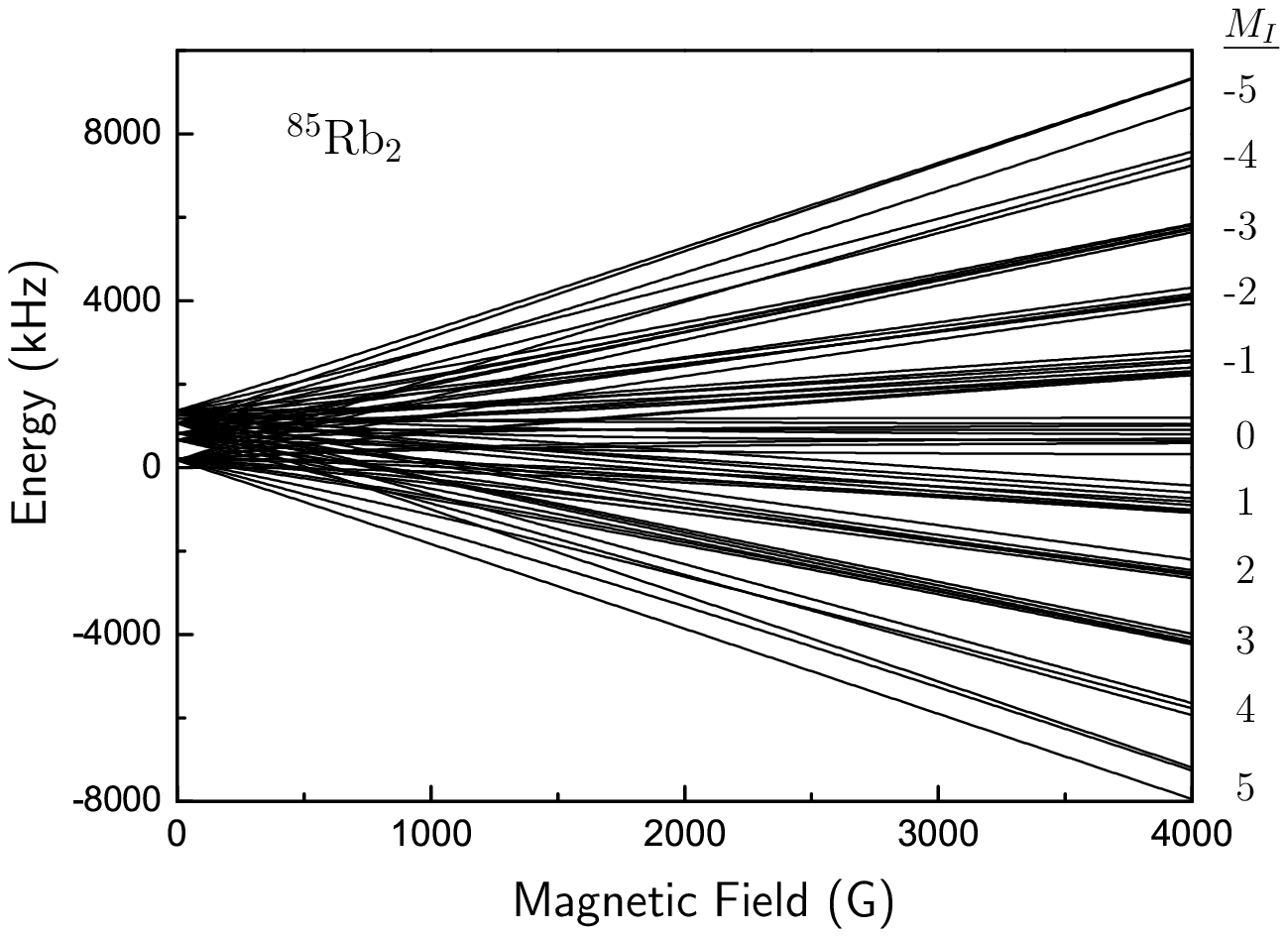}
\includegraphics[width=0.95\linewidth]{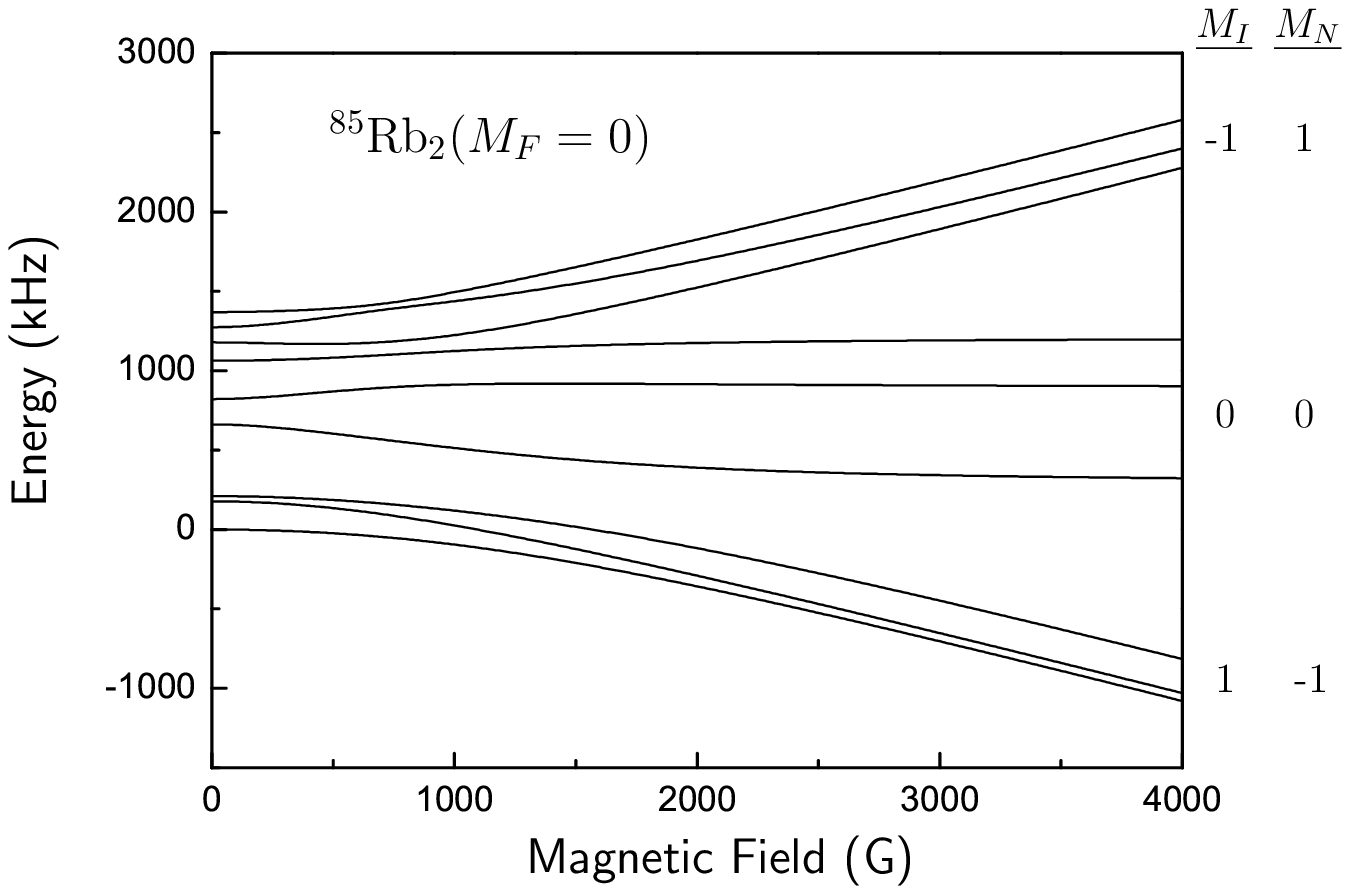}
\caption{\label{fig:04}%
Zeeman splitting for $N=1$ states of $^{85}$Rb$_2$. The
avoided crossings for $M_F=0$ are shown in the lower panel.}
\end{figure}

The quantum numbers that label the zero-field energy levels are
included in figure \ref{fig:02}. The total angular momentum
quantum number $F$ is always a good quantum number at zero
field. In some cases, when there is only one pair of values $I$
and $N$ that can couple to give the resultant $F$, $I$ is also
a good quantum number. Otherwise, $I$ is mixed and the values
given in figure \ref{fig:02} are ordered according to their
contribution to the eigenstate: the first quantum number listed
identifies the largest contribution.

The Zeeman splittings for $N=1$ states of $^{85}$Rb$_2$ and
$^{87}$Rb$_2$ for different ranges of magnetic fields are shown
in figures \ref{fig:03} and \ref{fig:04}. Each zero-field level
splits into $2F+1$ states with different projection quantum
numbers $M_{F}$. Although in principle both the nuclear
($H_{\rm IZ}$) and the rotational ($H_{\rm NZ}$) Zeeman terms
contribute to the splitting, $g_{\rm Rb} \gg g_{\rm r}$ so that
the rotational Zeeman term contributes only about 1\% for
$^{85}$Rb$_2$ and less than 0.5\% for $^{87}$Rb$_2$.

In contrast with the $N=0$ case, the Hamiltonian for $N=1$ is
not diagonal and energy levels corresponding to the same
$M_{F}$ value display avoided crossings. The magnetic field
values at which the avoided crossings are found, between 0 and
2000 G for $^{85}$Rb$_2$ (lower panel of figure \ref{fig:04})
and between 0 and 200 G for $^{87}$Rb$_2$ (lower panel of
figure \ref{fig:03}), scale with the ratio between the electric
quadrupole constant and the nuclear $g$-factor.

For larger magnetic fields, $M_{I}$ and $M_{N}$ become
individually good quantum numbers and the energy levels
corresponding to the same value of $M_{I}$ gather together.
Both features are illustrated in figure \ref{fig:04} where, for
the sake of clarity, the values of $M_{N}$ are included only in
the lower panel. Equation \ref{eq:HIZ1} shows that the matrix
representation of the nuclear Zeeman term in the spin-coupled
basis is diagonal with nonzero elements proportional to $M_{I}$
and independent of any other quantum number. As the magnetic
field increases the nuclear Zeeman terms becomes dominant and
the slope of the energy levels is determined by $M_{I}$.

The results in figure \ref{fig:04} neglect the diamagnetic
Zeeman interaction, which is not completely negligible at the
highest fields considered (up to 4000 G). The justification for
this is as follows. The Hamiltonian for the diamagnetic Zeeman
interaction \cite{Ramsey:1952} consists of two terms
proportional to the square of the magnetic field: one depending
on the trace of the magnetizability tensor and the other is
proportional to its anisotropic part. The first term has a
value around 200 kHz at 4000 G for $^{85}$Rb$_2$. Although this
quantity is not negligible, it has not been included because it
simply shifts all the energy levels by the same amount and has
no effect on splittings. The second term is diagonal in the
spin-coupled basis set and its nonzero elements depend on $N$
and $M_{N}$. For $^{85}$Rb$_2$ at 4000 G it would shift the
energy levels by about 15 kHz. It is therefore very small
compared to the nuclear Zeeman effect.

\section{Conclusions}
\label{sec:con} We have explored the hyperfine energy levels
and Zeeman splitting patterns for low-lying rovibrational
states of homonuclear alkali-metal dimers in $^1\Sigma$ states.
We have calculated the nuclear hyperfine coupling constants for
all common isotopic species of the homonuclear dimers from
Li$_2$ to Cs$_2$ and explored the energy level patterns in
detail for $^{85}$Rb$_2$ and $^{87}$Rb$_2$.

For rotationless molecules ($N=0$ states), the zero-field
splitting arises almost entirely from the scalar nuclear
spin-spin coupling. The levels are characterized by a total
nuclear spin quantum number $I$ and states with different
values of $I$ are separated by amounts between 90 Hz for
$^{41}$K$_2$ and 160 kHz for $^{133}$Cs$_2$. When a magnetic
field is applied, each level splits into $2I+1$ components but
all the levels with a particular value of $M_I$ are parallel.
This is different from the heteronuclear case, and for
homonuclear molecules $I$ remains a good quantum number in a
magnetic field. However, the projection quantum numbers
$M_{I1}$ and $M_{I2}$ for individual nuclei do not become
nearly good quantum numbers at high fields for homonuclear
molecules. These differences in quantum numbers may have
important consequences for spectroscopic selection rules and
for the collisional stability of molecules in excited hyperfine
states.

Molecules in excited rotational states are also of considerable
interest. In particular, molecules in $N=1$ states may be
collisionally stable because transitions between even and odd
rotational levels require a change in nuclear exchange
symmetry. Molecules in excited rotational states have
anisotropic quadrupole-quadrupole interactions that are
longer-range than dispersion interactions. The hyperfine energy
level patterns are considerably more complicated for $N=1$
states than for $N=0$ states and $I$ is not in general a good
quantum number even at zero field.

The results of the present paper will be important in studies that
produce ultracold molecules in low-lying rovibrational levels, where
it is important to understand and control the population of
molecules in different hyperfine states.

\section*{Acknowledgments}
The authors are grateful to EPSRC for funding of the
collaborative project QuDipMol under the ESF EUROCORES
Programme EuroQUAM and to the UK National Centre for
Computational Chemistry Software for computer facilities.

\begin{widetext}
\appendix*
\section{A} \label{sec:app}

Explicit expressions for the matrix elements of the molecular
Hamiltonian terms are now provided. The equations are valid for
homonuclear molecules.

The matrix elements for the rotational term ($H_{\rm rot}$) are
given by
\begin{eqnarray}
\label{eq:Hrot1} \langle N M_N(I_{\rm 1} I_{\rm 1}) I M_{I}|
H_{\rm rot} | N' M_{N'} (I_{\rm 1} I_{\rm 1}) I' M_{I'}
   \rangle & = &
\delta_{N N'}\,\delta_{M_NM_{N'}}\,\delta_{I I'}\,\delta_{M_{I} M_{I'}} B \, N(N+1) \\
\label{eq:Hrot2} \langle N (I_{\rm 1} I_{\rm 1}) I F M_{F}| H_{\rm sc}|
N' (I_{\rm 1} I_{\rm 1}) I' F' M_{F'}
   \rangle & = &
\delta_{N N'}\,\delta_{I I'}\,\delta_{F
F'}\,\delta_{M_{F} M_{\rm F'}} B\, N(N+1).
\end{eqnarray}

The matrix elements for the electric quadrupole interaction ($H_{\rm
Q}$) are given by
\begin{eqnarray}
\langle N M_N(I_{\rm 1} I_{\rm 1}) I M_{I}| &H_{\rm
Q}& | N' M_{N'} (I_{\rm 1} I_{\rm 1}) I' M_{I'}
   \rangle =
\frac{(eqQ)_{1}}{4}\,(-1)^{2I_{1}}\{4(2N+1)(2N'+1)
(2I+1)(2I'+1)\}^{1/2} \nonumber\\
&\times&  \left(\begin{array}{ccc}
N & 2 & N'  \\
0 & 0 & 0
\end{array} \right)
\left(\begin{array}{ccc}
I_{1} & 2 & I_{1}  \\
-I_{1} & 0 & I_{1}
\end{array} \right)^{-1}
\left\{\begin{array}{ccc}
I_{1} & I & I_{1}  \\
I'    & I_{1} & 2
\end{array} \right\} \nonumber\\
&\times&  \sum_{F M_{F}} \Bigg[(-1)^{F+2M_{F}} (2F+1)
\left(\begin{array}{ccc}
N & I & F  \\
M_N & M_{I} & -M_{F}
\end{array} \right) \nonumber\\
\label{eq:HQ1} &\times& \left(\begin{array}{ccc}
N' & I' & F  \\
M_{N'} & M_{I'} & -M_{F}
\end{array} \right)
\left\{\begin{array}{ccc}
I & N & F  \\
N'& I' & 2
\end{array} \right\}
\Bigg] \\
\langle N (I_{\rm 1} I_{\rm 1}) I F M_{F}| &H_{\rm Q}&| N'
(I_{\rm 1} I_{\rm 1}) I' F' M_{F'}
   \rangle =
\delta_{F F'}\,\delta_{M_{F} M_{\rm F'}}
\frac{(eqQ)_{1}}{4}\,(-1)^{N+N'+F+2I_{1}}
\{4(2N+1)(2N'+1)\}^{1/2} \nonumber \\
\label{eq:HQ2} &\times& \{(2I+1)(2I'+1)\}^{1/2}
\left(\begin{array}{ccc}
N & 2 & N'  \\
0 & 0 & 0
\end{array} \right)
\left(\begin{array}{ccc}
I_{1} & 2 & I_{1}  \\
-I_{1} & 0 & I_{1}
\end{array} \right)^{-1}
\left\{\begin{array}{ccc}
I & N & F  \\
N'& I'& 2
\end{array} \right\}
\left\{\begin{array}{ccc}
I_{1} & I & I_{1}  \\
I'    & I_{1} & 2
\end{array} \right\}
\end{eqnarray}
The matrix elements for the spin-rotation interaction ($H_{\rm
IJ}$) are given by
\begin{eqnarray}
\langle N M_N(I_{\rm 1} I_{\rm 1}) I M_{I}| & H_{\rm IJ} & |
N' M_{N'} (I_{\rm 1} I_{\rm 1}) I' M_{I'}
   \rangle = \delta_{N N'}\delta_{I I'}c_1(-1)^{N-M_N-M_{I}+2I_{1}+1}2(2I+1) \nonumber\\
& \times & \{(2N+1)N(N+1)(2I_{1}+1)I_{1}(I_{1}+1)\}^{1/2}
\left\{\begin{array}{ccc}
I_{1} & I & I_{1}  \\
I & I_{1} & 1
\end{array} \right\} \nonumber\\
\label{eq:IJ1} & \times & \sum_{p}\Bigg[ (-1)^{p}
\left(\begin{array}{ccc}
N & 1 & N  \\
-M_{N} & p & M_{N'}
\end{array} \right)
\left(\begin{array}{ccc}
I & 1 & I  \\
-M_{I} & -p & M_{I'}
\end{array} \right)
\Bigg] \\
\langle N (I_{\rm 1} I_{\rm 1}) I F M_{F}| & H_{\rm IJ} & | N'
(I_{\rm 1} I_{\rm 1}) I' F' M_{F'}
   \rangle = -2\delta_{N N'}\,\delta_{I I'}\,\delta_{F
F'}\,\delta_{M_{F} M_{\rm F'}} c_1 (-1)^{F+2I_{1}-M_{F}-I}
(2F+1) \nonumber\\
& \times & (2I+1)\{(2N+1)N(N+1)(2I_{1}+1)I_{1}(I_{1}+1)\}^{1/2}
\left\{\begin{array}{ccc}
I_{1} & I & I_{1}  \\
I & I_{1} & 1
\end{array} \right\} \nonumber\\
& \times & \sum_{F'',M_{\rm F''}} \Bigg[ (-1)^{F''+M_{\rm F''}}
(2F''+1) \left\{\begin{array}{ccc}
N & F & I  \\
F'' & N & 1
\end{array} \right\} \nonumber\\
\label{eq:IJ2} & \times & \sum_{p}\bigg[ (-1)^{p}
\left(\begin{array}{ccc}
F & 1 & F''  \\
-M_{F} & p & M_{\rm F''}
\end{array} \right)^{2}\bigg]
\left\{\begin{array}{ccc}
I & F'' & N  \\
F & I & 1
\end{array} \right\}
\Bigg]
\end{eqnarray}
The matrix elements for the scalar nuclear spin-spin
interaction ($H_{\rm sc}$) are given by
\begin{eqnarray}
\label{eq:Hesc1} \langle N M_N(I_{\rm 1} I_{\rm 1}) I M_{I}|
H_{\rm sc} | N' M_{N'} (I_{\rm 1} I_{\rm 1}) I' M_{I'}
   \rangle & = & \delta_{N N'}\delta_{M_NM_{N'}}\delta_{I I'}\delta_{M_{I}
M_{I'}} \frac{1}{2} c_4 [I(I+1)-2I_{\rm 1}(I_{\rm 1}+1)]\\
\label{eq:Hesc2} \langle N (I_{\rm 1} I_{\rm 1}) I F M_{F}| H_{\rm
sc}| N' (I_{\rm 1} I_{\rm 1}) I' F' M_{F'}
   \rangle & = & \delta_{N N'}\,\delta_{I I'}\,\delta_{F F'}\,\delta_{M_{F} M_{\rm
F'}} \frac{1}{2} c_4 [I(I+1)-2I_{\rm 1}(I_{\rm 1}+1)]
\end{eqnarray}
The matrix elements for the tensor nuclear spin-spin
interaction ($H_{\rm t}$) are given by
\begin{eqnarray}
\langle N M_N(I_{\rm 1} I_{\rm 1}) I M_{I}| & H_{\rm t} & | N'
M_{N'} (I_{\rm 1} I_{\rm 1}) I' M_{I'}
   \rangle =
-c_3 \sqrt{30} (-1)^{I-M_{I}-M_N} I_{1}(I_{1}+1)(2I_{1}+1) \{(2N+1)(2N'+1)\}^{1/2} \nonumber \\
& \times & \{(2I+1)(2I'+1)\}^{1/2} \left(\begin{array}{ccc}
N & 2 & N'  \\
0 & 0 & 0
\end{array} \right)
\left\{\begin{array}{ccc} I_{1} & I_{1} & 1  \\
I_{1} & I_{1} & 1   \\
I  & I'  & 2
\end{array} \right\} \nonumber \\
\label{eq:Ht1} & \times & \sum_{p}\Bigg[ (-1)^{p}
\left(\begin{array}{ccc}
N & 2 & N'  \\
-M_{N} & p & M_{N'}
\end{array} \right)
 \left(\begin{array}{ccc}
I & 2 & I'  \\
-M_{I} & -p & M_{I'}
\end{array} \right)
 \Bigg] \\
\langle N (I_{\rm 1} I_{\rm 1}) I F M_{F}| & H_{\rm t} & | N'
(I_{\rm 1} I_{\rm 1}) I' F' M_{F'}
   \rangle =
-\delta_{F F'}\,\delta_{M_{F} M_{\rm F'}} c_3
\sqrt{30} (-1)^{N'+N+I+F} I_{1}(I_{1}+1)(2I_{1}+1)  \nonumber \\
\label{eq:Ht2}  & \times & \{(2N+1)(2N'+1)(2I+1)(2I'+1)\}^{1/2}
 \left(\begin{array}{ccc}
N & 2 & N'  \\
0 & 0 & 0
\end{array} \right)
\left\{\begin{array}{ccc}
N & I & F  \\
I' & N' & 2
\end{array} \right\}
\left\{\begin{array}{ccc}
I_{1} & I_{1} & 1  \\
I_{1} & I_{1} & 1   \\
I  & I'  & 2
\end{array} \right\}
\end{eqnarray}
The matrix elements for the nuclear Zeeman term ($H_{\rm IZ}$)
are given by
\begin{eqnarray}
\langle N M_N(I_{\rm 1} I_{\rm 1}) I M_{I}| & H_{\rm IZ} & |
N' M_{N'} (I_{\rm 1} I_{\rm 1}) I' M_{I'} \rangle
= -\delta_{N N'}\delta_{M_NM_{N'}} \delta_{I I'}\delta_{M_{I} M_{I'}} g_{1}\mu_{\rm N} B_{\rm
Z}(1-\sigma_{1})M_{I} \label{eq:HIZ1} \\
\langle N (I_{\rm 1} I_{\rm 1}) I F M_{F}| & H_{\rm IZ} & | N'
(I_{\rm 1} I_{\rm 1}) I' F' M_{F'}
   \rangle =
-\delta_{N N'}\,\delta_{I I'}\,\delta_{M_{F}
M_{\rm F'}}g_{1}\mu_{\rm N}B_{\rm Z}(1-\sigma_{1})(-1)^{2F-M_{F}} \nonumber\\
& \times & (-1)^{N+2I_{1}}(2I+1)\{4(2F'+1)(2F+1)(2I_{1}+1)I_{1}(I_{1}+1)\}^{1/2} \nonumber\\
\label{eq:HIZ2} & \times &
\left(\begin{array}{ccc}
F & 1 & F'  \\
-M_{F} & 0 & M_{F}
\end{array} \right)
\left\{\begin{array}{ccc}
I & F & N  \\
F' & I & 1
\end{array} \right\}
\left\{\begin{array}{ccc}
I_{1} & I & I_{1}  \\
I & I_{1} & 1
\end{array} \right\}
\end{eqnarray}
The matrix elements for the rotational Zeeman effect ($H_{\rm
NZ}$) are given by
\begin{eqnarray}
\label{eq:HNZ1} \langle N M_N(I_{\rm 1} I_{\rm 1}) I M_{I}| &
H_{\rm NZ} & | N' M_{N'} (I_{\rm 1} I_{\rm 1}) I' M_{I'}
\rangle = -\delta_{N N'}\delta_{M_NM_{N'}}\delta_{I
I'}\delta_{M_{I} M_{I'}} g_{\rm r}\mu_{\rm N}B_{\rm Z}M_N \\
\langle N (I_{\rm 1} I_{\rm 1}) I F M_{F}| & H_{\rm NZ} & | N'
(I_{\rm 1} I_{\rm 1}) I' F' M_{F'}
   \rangle = -\delta_{N N'}\,\delta_{I I'}\,\delta_{M_{F} M_{\rm
F'}}g_{\rm r}\mu_{\rm N}B_{\rm Z}(-1)^{N+I+F+F'-M_{F}+1} \nonumber\\
\label{eq:HNZ2} & \times & \{(2F'+1)(2F+1)(2N+1)N(N+1)\}^{1/2}
\left(\begin{array}{ccc}
F & 1 & F'  \\
-M_{F} & 0 & M_{F}
\end{array} \right)
\left\{\begin{array}{ccc}
N & F & I  \\
F' & N & 1
\end{array} \right\}
\end{eqnarray}
\end{widetext}


\begin{thebibliography}{31}

\bibitem{Ospelkaus:2008}
S.~Ospelkaus, A.~Pe'er, K.K. Ni, J.J. Zirbel, B.~Neyenhuis,
S.~Kotochigova,
  P.S. Julienne, J.~Ye, D.S. Jin, Nature Physics \textbf{4}, 622 (2008)

\bibitem{Ni:KRb:2008}
K.K. Ni, S.~Ospelkaus, M.H.G. {de Miranda}, A.~Pe'er, B.~Neyenhuis,
J.J.
  Zirbel, S.~Kotochigova, P.S. Julienne, D.S. Jin, J.~Ye,
  arXiv:quant-ph/0808.2963  (2008)

\bibitem{Danzl:v73:2008}
J.G. Danzl, E.~Haller, M.~Gustavsson, M.J. Mark, R.~Hart,
N.~Bouloufa,
  O.~Dulieu, H.~Ritsch, H.C. N\"agerl, Science \textbf{321}, 1062 (2008)

\bibitem{Danzl:ground:2008}
J.G. Danzl, M.J. Mark, E.~Haller, M.~Gustavsson, R.~Hart, H.C.
N\"agerl,
  \emph{to be published} (2008)

\bibitem{Lang:cruising:2008}
F.~Lang, P.~{van der Straten}, B.~Brandst\"atter, G.~Thalhammer,
K.~Winkler,
  P.S. Julienne, R.~Grimm, J.~{Hecker Denschlag}, Nature Phys. \textbf{4}, 223
  (2008)

\bibitem{Lang:ground:2008}
F.~Lang, K.~Winkler, C.~Strauss, R.~Grimm, J.H. Denschlag,
  arXiv:quant-ph/0809.0061  (2008)

\bibitem{Sage:2005}
J.M. Sage, S.~Sainis, T.~Bergeman, D.~DeMille, Phys. Rev. Lett.
  \textbf{94}(20), 203001 (2005)

\bibitem{Hudson:PRL:2008}
E.R. Hudson, N.B. Gilfoy, S.~Kotochigova, J.M. Sage, D.~DeMille,
Phys. Rev.
  Lett. \textbf{100}, 203201 (2008),
  \texttt{http://link.aps.org/abstract/PRL/v100/e203201}

\bibitem{Viteau:2008}
M.~Viteau, A.~Chotia, M.~Allegrini, N.~Bouloufa, O.~Dulieu,
D.~Comparat,
  P.~Pillet, Science \textbf{321}, 232 (2008)

\bibitem{Deiglmayr:2008}
J.~Deiglmayr, A.~Grochola, M.~Repp, K.~M\"ortlbauer, C.~Gl\"uck,
J.~Lange,
  O.~Dulieu, R.~Wester, M.~Weidem\"uller, arXiv:quant-ph/0807.3272  (2008)

\bibitem{Aldegunde:polar:2008}
J.~Aldegunde, B.A. Rivington, P.S. \.Zuchowski, J.M. Hutson, Phys.
Rev. A
  \textbf{78}, 033434 (2008)

\bibitem{Ramsey:1952}
N.F. Ramsey, Phys. Rev. \textbf{85}, 60 (1952)

\bibitem{Brown}
J.M. Brown, A.~Carrington, \emph{Rotational Spectroscopy of Diatomic
Molecules}
  (Cambridge University Press, Cambridge, 2003)

\bibitem{Bryce:2003}
D.L. Bryce, R.E. Wasylishen, Acc. Chem. Res. \textbf{36}, 327 (2003)

\bibitem{Stone:2005}
N.J. Stone, At. Data Nucl. Data Tables \textbf{90}, 75 (2005)

\bibitem{Brooks:1963}
R.A. Brooks, C.H. Anderson, N.F. Ramsey, Phys. Rev. Letters
\textbf{10}, 441
  (1963)

\bibitem{Esbroeck:1985}
P.E. Van~Esbroeck, R.A. McLean, T.D. Gaily, R.A. Holt, S.D. Rosner,
Phys. Rev.
  A \textbf{32}, 2595 (1985)

\bibitem{ADF1}
G.~te~Velde, F.M. Bickelhaupt, S.J.A. van Gisbergen,
C.~Fonseca~Guerra, E.J.
  Baerends, J.G. Snijders, T.~Ziegler, J. Comput. Chem. \textbf{22}, 931 (2001)

\bibitem{ADF3}
\emph{{ADF2007.01}}, http://www.scm.com (2007), {SCM}, {T}heoretical
  {C}hemistry{,} {V}rije {U}niversiteit{,} {A}msterdam{,} {T}he {N}etherlands

\bibitem{Hessel:1979}
M.M. Hessel, C.R. Vidal, J. Chem. Phys. \textbf{70}, 4439 (1979)

\bibitem{Kusch:1978}
P.~Kusch, M.M. Hessel, J. Chem. Phys. \textbf{68}, 2591 (1978)

\bibitem{Engelke:1984}
F.~Engelke, H.~Hage, U.~Sch\"uhle, Chem. Phys. Lett. \textbf{106},
535 (1984)

\bibitem{Amiot:1985}
C.~Amiot, P.~Crozet, J.~Verg\`es, Chem. Phys. Lett. \textbf{121},
390 (1985)

\bibitem{Raab:1982}
M.~Raab, G.~H\"oning, W.~Demtr\"oder, C.R. Vidal, J. Chem. Phys.
\textbf{76},
  4370 (1982)

\bibitem{Zare}
R.N. Zare, \emph{Angular Momentum} (John Wiley \& Sons, 1987)

\bibitem{Soldan:2002}
P.~Sold\'{a}n, M.T. Cvita\v{s}, J.M. Hutson, P.~Honvault, J.M.
Launay, Phys.
  Rev. Lett. \textbf{89}(15), 153201 (2002)

\bibitem{Cvitas:bosefermi:2005}
M.T. Cvita\v{s}, P.~Sold\'{a}n, J.M. Hutson, P.~Honvault, J.M.
Launay, Phys.
  Rev. Lett. \textbf{94}(3), 033201 (2005)

\bibitem{Cvitas:hetero:2005}
M.T. Cvita\v{s}, P.~Sold\'{a}n, J.M. Hutson, P.~Honvault, J.M.
Launay, Phys.
  Rev. Lett. \textbf{94}(20), 200402 (2005)

\bibitem{Quemener:2005}
G.~Qu\'{e}m\'{e}ner, P.~Honvault, J.M. Launay, P.~Sold\'{a}n, D.E.
Potter, J.M.
  Hutson, Phys. Rev. A \textbf{71}(3), 032722 (2005)

\bibitem{Cvitas:li3:2007}
M.T. Cvita\v{s}, P.~Sold\'{a}n, J.M. Hutson, P.~Honvault, J.M.
Launay, J. Chem.
  Phys. \textbf{127}, 074302 (2007)

\bibitem{Hutson:IRPC:2007}
J.M. Hutson, P.~Sold\'{a}n, Int. Rev. Phys. Chem. \textbf{26}(1), 1
(2007)

\end{thebibliography}
\end{document}